\documentclass[mathpazo]{cicp}
\usepackage{graphicx}%
\usepackage{dcolumn}%
\usepackage{bm}%
\usepackage[utf8]{inputenc}
\usepackage[T1]{fontenc}
\usepackage{threeparttable}
\usepackage[abbreviations]{glossaries-extra}
\usepackage{mathptmx}
\usepackage{etoolbox}
\usepackage{color}
\usepackage{CJKutf8}
\usepackage{xcolor}
\usepackage{soul}
\usepackage{float}
\usepackage{multirow}
\usepackage{colortbl}
\usepackage{booktabs}

\soulregister\cite7
\soulregister\ref7
\soulregister\eqref7
\soulregister\pageref7

\begin{document}
\title{Mandelic acid single-crystal growth: Experiments vs numerical simulations}

\author[Tan Q. et~al.]{Q. Tan\affil{1},
      S.A.~Hosseini\affil{1,2}\comma\corrauth, A.~Seidel-Morgenstern\affil{3}, D.~Th\'evenin\affil{1}, H.~Lorenz\affil{3}}
      
\address{\affilnum{1}\ Laboratory of Fluid Dynamics and Technical Flows, University of Magdeburg ``Otto von Guericke'', D-39106 Magdeburg, Germany. \\
          \affilnum{2}\ Department of Mechanical and Process Engineering, ETH Z\"urich, 8092 Z\"urich, Switzerland.\\
          \affilnum{3}\ Max Planck Institute for Dynamics of Complex Technical Systems (MPI DCTS), 39106 Magdeburg, Germany.}
          
\emails{{\tt seyed.hosseini@ovgu.de} (S.A.~Hosseini)}

\begin{abstract}
Mandelic acid is an enantiomer of interest in many areas, in particular for the pharmaceutical industry. One of the approaches to produce enantiopure mandelic acid is through crystallization from an aqueous solution. We propose in this study a numerical tool based on lattice Boltzmann simulations to model crystallization dynamics of (S)-mandelic acid. The solver is first validated against experimental data. It is then used to perform parametric studies concerning the effects of important parameters such as supersaturation and seed size on the growth rate. It is finally extended to investigate the impact of forced convection on the crystal habits. Based on there parametric studies, a modification of the reactor geometry is proposed that should reduce the observed deviations from symmetrical growth with a five-fold habit.
\end{abstract}


\maketitle

\section{Introduction}
\label{sec1}
\newabbreviation{lbm}{LBM}{lattice Boltzmann method}
\newabbreviation{lb}{LB}{lattice Boltzmann}
\newabbreviation{ns}{NS}{Navier-Stokes}
\newabbreviation{pde}{PDE}{partial differential equation}
Mandelic acid is an aromatic alpha-hydroxy acid, with formula ${\rm C}_8{\rm H}_8{\rm O}_3$. It is a white crystalline powder that is soluble in water and most common organic solvents. It has a density of 1.3 g/cm$^3$  and molecular weight of 152.5 g/mol. It is particularly important in the pharmaceutical industry for the organic synthesis of pharmaceutical components. For instance an ester of mandelic acid is essential to produce homatropine, used in eye drops as both a cycloplegic and mydriatic substance. In addition, it is also popular in the production of face peeling components~\cite{taylor1999summary}, urinary tract infection treatments~\cite{brittain2002mandelic}, and for oral antibiotics~\cite{sharon2018mandelic}. In toxicological studies, the concentration of styrene or styrene oxide is quantified by converting it into mandelic acid.
\begin{figure}[H]           
\centering	
	\includegraphics[width= 0.4\textwidth]{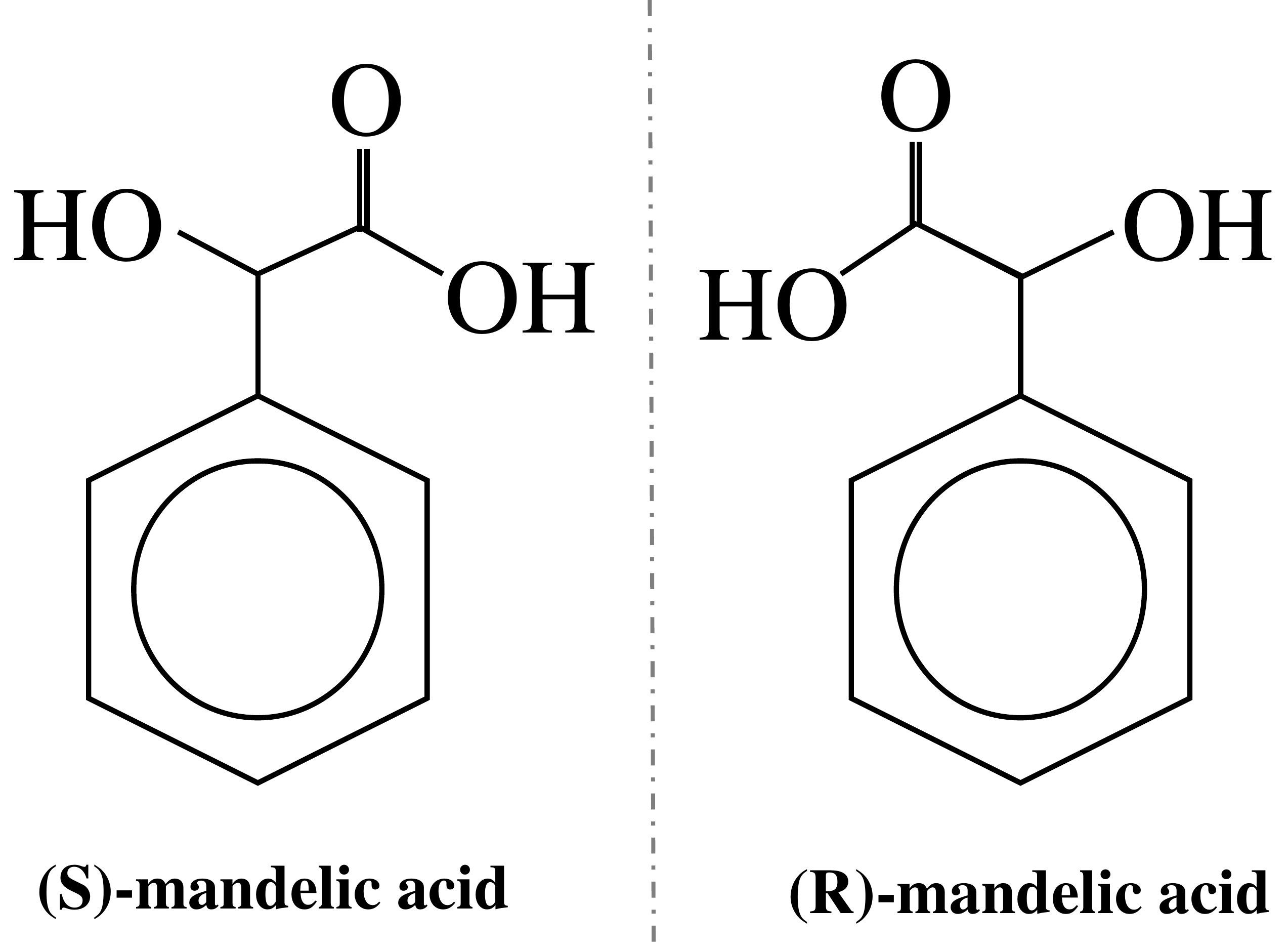}	
	\caption{Molecular structure of mandelic acid enantiomers. }
	\label{aass}
\end{figure}

Mandelic acid exists in two enantiomeric forms as shown in Fig.~\ref{aass}, (S)- and (R)- mandelic acid. Most practical applications require the enantiopure form~\cite{brittain2002mandelic}. Amongst the different approaches to separate enantiomers, crystallization process such as classical resolution and preferential crystallization approaches are frequently used~\cite{lorenz2014processes}. In such separation processes, the properties of the crystalline products such as crystal size and shape are largely determined by the growth process, which in turn depends on the crystallization conditions. 
In the pharmaceutical industry, the resulting crystal morphology is often of great importance, since it influences the rate of dissolution and the absorption of drugs. Compressibility, hardness, and flow properties of the drug are also strongly dependent on the crystal form~\cite{higgins1994numerical}.
Accurate investigations regarding crystal growth are difficult because the growth process varies greatly even under similar conditions: \emph{crystal growth dispersion} is the term used to describe the fact that crystals, although initially of same shape and size, can rapidly grow differently even under the same growth conditions~\cite{srisanga2015crystal,ma2008crystal}. The main reason for these growth differences is probably related to minute tensions and deformations, leading in turn to minimal structural differences~\cite{hofmann2004kristallisation}. Other reasons are accidental deposits, or deposits of foreign bodies on the growing crystals' surface, which lead to incorporation into the crystal and ultimately different growth.
A proper understanding of growth conditions and their effect on the final product is therefore essential to design and scale-up production units for enantiopure substances.\\
A lot of experimental studies have been conducted concerning crystallization-based enantioseparation process including growth kinetics of mandelic acid, e.g.~\cite{alvarez2004online,lorenz2014processes, coquerel2006preferential, gansch2021continuous, gou2012investigation, perlberg2005crystal, srisanga2015crystal,codan2013growth}. However, numerical studies regarding crystal habit and size of enantiopure S-mandelic acid remain scarce. The phase-field  method has been shown in general to be a powerful tool for modeling structural evolution of materials and crystals. It is now widely used for modeling solidification~\cite{boettinger2002phase,nestler2002phase} and grain growth~\cite{chen1994computer,takaki2016two,karma1998quantitative,tourret2017grain}. The phase-field approach has also been used in the context of the lattice Boltzmann method, now widely recognized as an efficient alternative to classical tools, to simulate solidification processes~\cite{younsi2016anisotropy,lin2014three,wang2019brief,rojas2015phase,schiedung2020simulation,vakili2020multi,m2019non}. This approach can reproduce numerically the solid-liquid interface interactions and the hydrodynamic effects affecting the habits of growing crystals~\cite{medvedev2005lattice,Henniges2017,medvedev2006influence,sakane2018three,chakraborty2007enthalpy,tan2021modeling}.\\
In this contribution, we study the growth of a single (S)-mandelic acid crystal under different conditions (supersaturation, initial crystal size, flow rate) with a previously developed and validated lattice Boltzmann-based numerical model~\cite{tan2021modeling}. All simulations presented in this article are carried out using the in-house solver ALBORZ~\cite{hosseini2020development}. The obtained results are validated and compared with experimental data. After validating the numerical procedure in a standalone manner via a self-convergence test, it is used to model the growth of a single (S)-mandelic acid rhombic seed at temperature and supersaturation corresponding to experimental settings; this provides a further, independent validation of the numerical model. The solver is then used to investigate the effect of different parameters such as supersaturation and initial seed size on crystal growth. Finally, a detailed study of the interaction between forced convection and crystal growth is presented. Analyzing the results, a simple solution is proposed to improve symmetrical growth under natural convection in the single-crystal cell used for all experimental investigations.
\section{Numerical method}
\subsection{Diffuse-interface formulation: governing equations}
In the phase-field method solid growth dynamics are expressed via a non-dimensional order parameter, $\phi$, going from (+1) in the solid to (-1) in the pure liquid phase. The space/time evolution equations are written as~\cite{jeong2001phase,beckermann1999modeling}:
\begin{multline}
    \tau_0 a_s^2(\textbf{n}) \frac{\partial \phi}{\partial t} = W_0^2  \bm{\nabla} \cdot \left(a_s^2(\textbf{n})\right) \bm{\nabla} \phi +  W_0^2 \bm{\nabla} \cdot \left (|\bm{\nabla} \phi|^2  \frac{\partial[a(\textbf{n})^2]}{\partial \bm{\nabla} \phi}\right )\\
    + (\phi - \phi^3) + \lambda U (1 - \phi^2)^2 ,
  \label{a}
\end{multline}
and:
\begin{equation}
    \frac{\partial U}{\partial t} + \left( \frac{1-\phi}{2}\right) \bm{u} \cdot \bm{\nabla} U  =D \bm{\nabla} \cdot \left(q(\phi) \bm{\nabla} U \right) - \frac{\partial \phi}{\partial t}.
     \label{b}
\end{equation}
Here $\tau = \tau_0 a_s^2(\textbf{n})$. The coefficient $\lambda$ describes the strength of the coupling between the phase-field and the supersaturation field, $U$. The parameter $\tau_0$ denotes the characteristic time and $W_0$ the characteristic width of the diffuse interfaces. In Eq.~(\ref{a}), the latent heat of melting is written as $L$. The specific heat capacity $c_p$ is assumed to be the same in the two phases (symmetric model). The quantity $\textbf{n} = - \frac{\bm{\nabla} \phi}{\left| \bm{\nabla} \phi \right|}$ is the unit vector normal to the crystal interface -- pointing from solid to fluid, while $a_s(\textbf{n})$ is the surface tension anisotropy function. In the context of the hexagonal mandelic acid crystal growth, this quantity is defined as:
\begin{equation}
    a_s(\textbf{n}) = 1 + \epsilon_{xy} \cos(6 \theta),
\end{equation}
\newabbreviation{rhs}{RHS}{right hand side}
\newabbreviation{lhs}{LHS}{left hand side}
where $\theta = \arctan(n_y/n_x)$. The numerical parameter $\epsilon_{xy}$ characterizes the anisotropy strength, and is set in the present study to $\epsilon_{xy} = 0.05$~\cite{karma1996phase}. The term $(\phi - \phi^3)$ is the derivative of the double-well potential. The last term in Eq.~(\ref{a}) is a source term accounting for the coupling between supersaturation $U$ and order parameter $\phi$. There, $(1 - \phi^2)^2$ is an interpolation function minimizing the bulk potential at $\phi = \pm 1$.\\
In Eq.~(\ref{b}), $\bm{u}$ denotes the local fluid velocity while $q(\phi) = (1 - \phi)$ is a function canceling out diffusion within the solid. As a consequence, solute transport is assumed to take place only within the fluid phase (one-sided model). The parameter $D$ is the diffusion coefficient of S-mandelic acid in water.
\subsection{Lattice Boltzmann formulation}
\paragraph{Flow field solver}
The flow field behavior, described by the incompressible \gls{ns} and continuity equations, is modeled using the classical isothermal \gls{lb} formulation consisting of the now-famous stream-collide operators:
\begin{equation}
    f_\alpha \left( \bm{r}+\bm{c}_\alpha \delta_t, t+\delta_t\right) - f_\alpha \left( \bm{r}, t\right) = \delta_t \Omega_\alpha\left( \bm{r}, t\right) + \delta_t F_{\alpha},
\end{equation}
where $f_\alpha$ and $\bm{c}_\alpha$ are the discrete populations and corresponding particle velocity vectors, $\bm{r}$ and $t$ are the position in physical space and time, $\delta_t$ is the time-step size. $F_\alpha$ represents contributions from external body forces defined as~\hbox{\cite{guo2002discrete}}:
\begin{equation}
    F_\alpha = w_\alpha \left(1-\frac{\delta_ t}{2\tau_f}\right)\left[\frac{\bm{F}\cdot\bm{c}_\alpha}{c_s^2} + \frac{\left(\bm{u}\otimes\bm{F}+\bm{u}\otimes\bm{F}^\dagger\right)\left(\bm{c}_\alpha\otimes\bm{c}_\alpha - c_s^2\bm{I}\right)}{2c_s^4}\right].
\end{equation}
{Here $\bm{F}$ is the external body force vector, $\bm{u}$ is the local fluid velocity vector, $\bm{I}$ is the unit rank-two tensor, $c_s$ is the so-called lattice sound speed, $w_\alpha$ are weights associated to each discrete velocity and $\tau_f$ is the relaxation time. Here, the external body force also includes interaction with the solid phase} as~\cite{beckermann1999modeling}:
\begin{equation}
  \bm{F}= -\frac{h\tau_f (1+\phi)^2 (1-\phi)\bm{u}}{4W_0^2},
\end{equation}
{where $\phi$ is the phase indicator detailed in the next paragraphs, $W_0$ is the interface thickness tied to the phase-field solver and} $h$ is a dimensionless constant, chosen as $h = 2.757$ following~\cite{beckermann1999modeling}. Due to the absence of fluid velocity within the solid crystal, the fluid velocity $\bm{u}$ is updated as:
\begin{equation}
    \bm{u^*} = \frac{(1-\phi)}{2}\bm{u},
\end{equation}
where the re-defined fluid velocity $\bm{u^*}$ is used in the equilibrium distribution function~\cite{beckermann1999modeling}. The collision operator $\Omega_\alpha$ follows the linear Bhatnagar-Gross-Krook approximation:
\begin{equation}
    \Omega_\alpha = \frac{1}{\tau_f}\left[f^{(eq)}_\alpha  - f_\alpha\right],
\end{equation}
\newabbreviation{edf}{EDF}{equilibrium distribution function}
where $f_\alpha^{(eq)}$ is the discrete isothermal \gls{edf} defined as:
\begin{equation}
    f_\alpha^{(eq)} = \rho w_\alpha\sum_i \frac{1}{i! c_s^{2i}} a^{(eq)}_i(\bm{u}):\mathcal{H}_{i}(\bm{c}_\alpha),
\end{equation}
where $a^{(eq)}_i$ and $\mathcal{H}_{i}(\bm{c}_\alpha)$ are the corresponding multivariate Hermite coefficients and polynomials of order $i$~\cite{hosseini2020compressibility}. Further information on the expansion along with detailed expressions of the \gls{edf} can be found in~\cite{shan2006kinetic,hosseini2019extensive,hosseini2020development}. In the present work, an extended range of stability is obtained by using a central Hermite multiple relaxation time implementation; corresponding details can be found in~\cite{hosseini2021central}. The relaxation time {$\tau_f$} is tied to the fluid kinematic viscosity{, $\nu$,} as:
\begin{equation}
    \tau_f = \frac{\nu}{c_s^2} + \frac{\delta_t}{2}.
\end{equation}
It must be noted that conserved variables, {i.e.}, density and momentum are defined as moments of the discrete distribution function:
\begin{equation}
    \rho = \sum_\alpha f_\alpha,
\end{equation}
\begin{equation}
    \rho \bm{u} = \sum_\alpha \bm{c}_\alpha f_\alpha.
\end{equation}
\paragraph{Advection-diffusion-reaction solver for supersaturation field}
The space/time evolution equation of the supersaturation field $U$ is modeled using an advection-diffusion-reaction \gls{lb}-based discrete kinetic equation. It is defined as\cite{ponce1993lattice,hosseini2020weakly,hosseini2019lattice}:
\begin{equation}
    g_\alpha \left( \bm{r}+\bm{c}_\alpha \delta_t, t+\delta_t\right) - g_\alpha \left( \bm{r}, t\right) = \delta_t \Omega_\alpha\left( \bm{r}, t\right) + \delta_t \dot{\omega}_\alpha,
\end{equation}
where {$g_\alpha$ are the corresponding discrete populations and} $\dot{\omega}_\alpha$ is the source term:
\begin{equation}
    \dot{\omega}_\alpha = - w_\alpha \frac{\partial \phi}{\partial t}.
\end{equation}
The collision operator $\Omega_\alpha$ for the supersaturation field is:
\begin{equation}
    \Omega_\alpha = \frac{1}{\tau_U}\left[g^{(eq)}_\alpha  - g_\alpha\right].
\end{equation}
where $g_\alpha^{(eq)}$ is the \gls{edf} defined as:
\begin{equation}
    g^{(eq)}_\alpha = w_\alpha U\left[ 1 + \frac{\bm{c}_\alpha \cdot \bm{u} }{c_s^2} \right].
\end{equation}
The supersaturation is the zeroth-order moment of $g_\alpha$:
\begin{equation}
    U = \sum_\alpha g_\alpha,
\end{equation}
and the relaxation coefficient is tied to the diffusion coefficient of mandelic acid {in the aqueous solution, $D$, as:}
\begin{equation}
    \tau_U = \frac{D q(\phi)}{c_s^2} + \frac{\delta_t}{2}.
\end{equation}
\paragraph{Solver for phase-field equation}
The phase-field equation is modeled using a modified \gls{lb} scheme defined as~\cite{walsh2010macroscale,cartalade2016lattice}:
\begin{multline}
  a_s^2(\bm{n}) h_\alpha(\bm{r} + \bm{c}_\alpha \delta_t, t + \delta_t) = h_\alpha(\bm{r},t) -  \left( 1 - a_s^2(\bm{n})  \right ) h_\alpha(\bm{r} + \bm{c}_\alpha \delta_t, t) - \\
   \frac{1}{\tau_\phi (\bm{r},t) }
  \left [ h_\alpha(\bm{r},t) - h_\alpha^{eq}(\bm{r},t) \right]  + w_\alpha Q_\phi (\bm{r},t)\frac{\delta_t}{\tau_0},
  \label{d}
\end{multline}
where the scalar function $Q_\phi$ is the source term of the phase-field defined as:
\begin{equation}
    Q_\alpha = (\phi - \phi^3) + \lambda U (1 - \phi^2)^2,
\end{equation}
while the \gls{edf}, $h_\alpha^{eq}$, is defined as:
\begin{equation}
    h_\alpha^{eq} = w_\alpha \left( \phi - \frac{1}{c_s^2} \bm{c}_\alpha \cdot \frac{W_0^2}{\tau_0} |\bm{\nabla} \phi|^2 \frac{\partial (a_s(\bm{n})^2}{\partial \bm{\nabla} \phi} \frac{\delta_t}{\delta_r} \right).
    \label{e}
\end{equation}
{where $\delta_r$ is the grid-size.} The local value of the order parameter $\phi$ is computed as:
\begin{equation}
    \phi = \sum_{\alpha} h_\alpha,
\end{equation}
while the relaxation is set to:
\begin{equation}
  \tau_\phi = \frac{1}{c_s^2}a_s^2(\bm{n})\frac{W_0^2}{\tau_0} + \frac{\delta_t}{2}.
\end{equation}
\section{Experimental setup}
\begin{figure}[H]           
\centering	
	\includegraphics[width=0.8\textwidth]{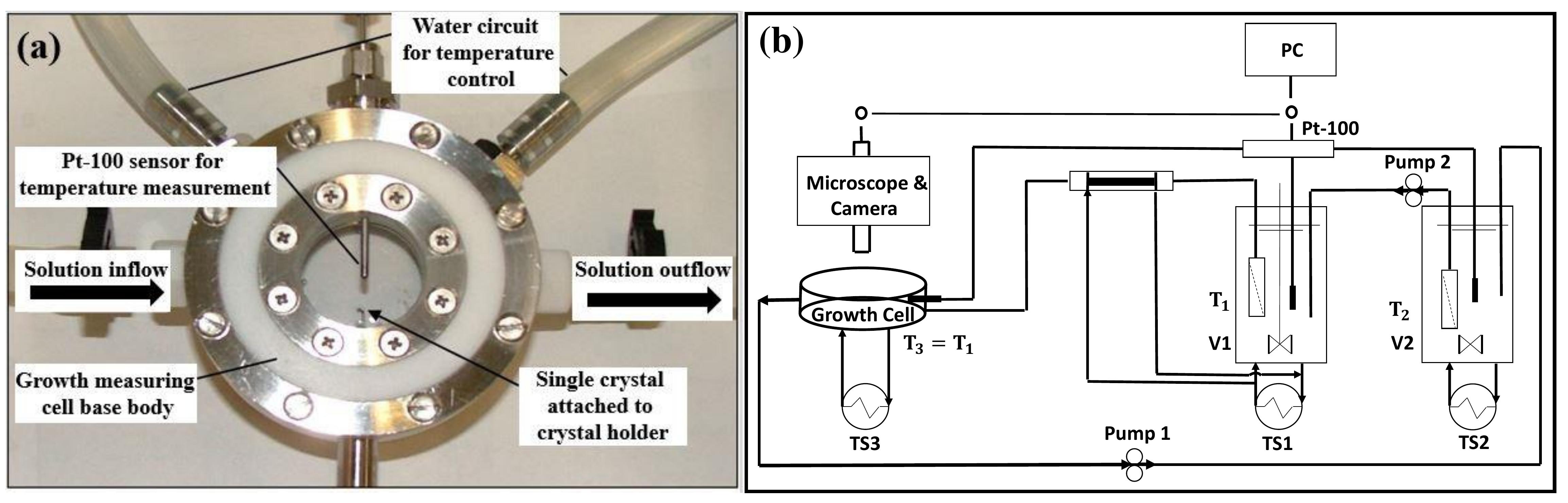}	
	\caption{Single-crystal growth cell used for all experiments: (a) photograph; (b) Schematic diagram of experimental arrangement for the measurement of growth rate of a single crystal~\cite{gou2012investigation, Juan}}.
	\label{ba}
\end{figure}

Experimental data for the growth rates have been obtained in the single-crystal growth cell~\cite{gou2012investigation, Juan} illustrated in Fig.~\ref{ba}. The supersaturated aqueous solution of mandelic acid is pumped into a constant-temperature cylindrical crystallization cell, with solution temperatures varying between 20 and 30 $^{\circ}$C. The temperature within the cell is maintained constant via a water-based cooling/heating system connected to a Pt-100 sensor monitoring the temperature inside the cell. Vessel 2, denoted V2 in Fig.~\ref{ba}b contains a saturated solution at temperature $T_2$ while vessel 1 (V1) was set to a lower temperature $T_1$, corresponding to the temperature of the cell. To create the supersaturated solution, the initially saturated solution in V2 is pumped into V1 and cooled down to $T_1$ before entering the growth cell. This effectively allows to control the supersaturation level of the incoming solution by choosing temperature $T_1$. In the present case, the supersaturation is defined as~\cite{mullin2001crystallization}:

\begin{equation}
    U = \frac{C_{sat,2} - C_{sat,1}}{C_{sat,1}}
\end{equation}

To start the experiment, the supersaturated solution is continuously pumped from vessel 1 to the growth cell, in which a single (S)-mandelic crystal is glued on the pin head of a crystal holder. Then, the solution is recycled to vessel 2 and the concentration of the solution is compensated. In that way, a stable degree of supersaturation is guaranteed during the whole process. A microscope with camera (Stemi2000C, company Carl Zeiss) is used to take pictures of the single crystal at every one hour. The images are afterwards post-processed by applying Carl Zeiss' Axio Vision software~\cite{gou2012investigation}. A picture of the single-crystal cell is shown in Fig.~\ref{ba}.\\

\section{Simulations and analysis of the results}
\label{sec3}
\subsection{Validation}
\subsubsection{Self-convergence of the numerical solver}
Based on the experiments, the evolution of mandelic acid without enantiomers follows habits with hexagonal symmetry. First, before going into further validation steps against experimental results, we look into the convergence behavior of the numerical scheme. To that end growth simulations are conducted using the hexagonal anisotropy function that will be used for the remainder of this work, starting with a rhombic initial seed. The seed is placed at the center of a fully periodic rectangular domain, with a length of 31 and a width of 26 mm. The perimeter of the initial rhombic crystal is $6.9$ mm and the initial supersaturation is set to $U = 0.06$. Simulations are conducted using four different spatial resolutions, $\delta_r\in\{0.04, 0.025, 0.02, 0.0125\}$ mm. Since the overall size of the numerical domain is kept fixed, an improved spatial resolution automatically comes with a larger number of grid points.

\begin{figure}[H]                        
\centering	
	\includegraphics[width=0.8\textwidth]{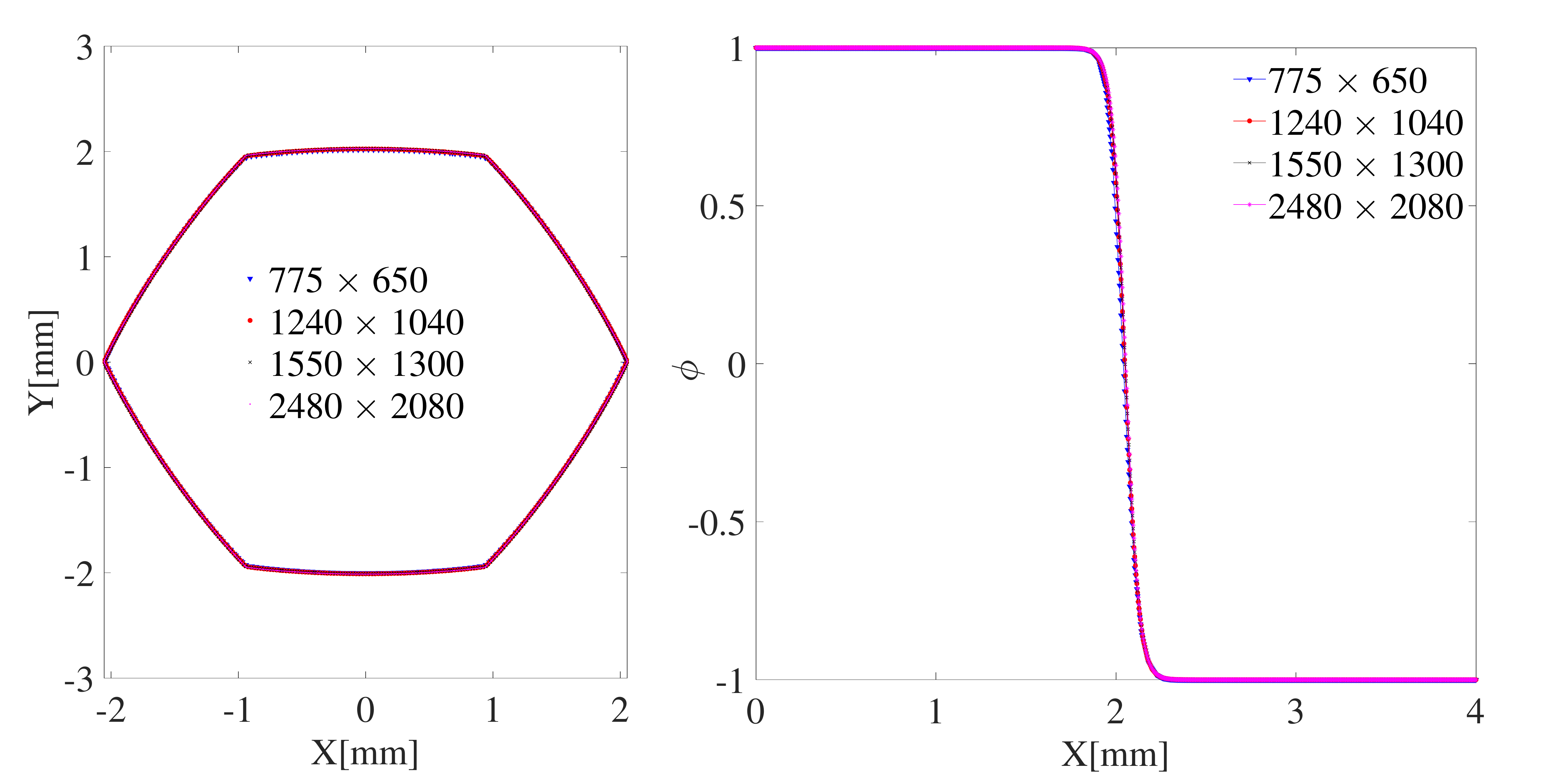}
	\caption{Left: $\phi = 0$ iso-contour, showing the boundary of the solid crystal (only the central part of the numerical domain is shown) after 16 hours; Right: Evolution of function $\phi$ in space along the line joining the center of the domain at (0,0) and point (4 mm,0) for $U = 0.06$ at increasing spatial resolution (grids with 775 $\times$ 650, 1240 $\times$ 1040, 1550 $\times$ 1300, and 2480 $\times$ 2080 points, respectively). }
	\label{cc}
\end{figure}

\begin{figure}[H]                        
\centering	
	\includegraphics[width=0.4\textwidth]{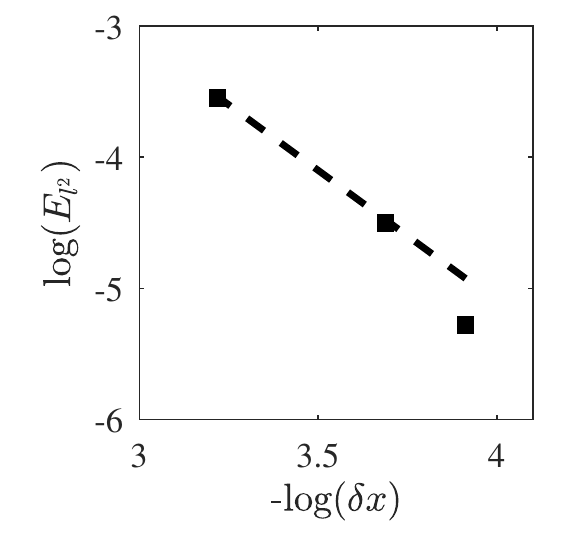}	
	\caption{Scaling of the $l^2$ norm of errors as obtained from the self-convergence study. Black markers represent error data from the simulations while the black dashed line shows the theoretical $-2$ slope.}
	\label{csfm}
\end{figure}

The highest resolution simulation with 2480 $\times$ 2080 points is used as reference to compute relative errors at the three lower spatial resolutions. The $\mathit{l^2}$ relative error norm is calculated based on the $\phi$-profiles plotted along the $x$-direction on the centerline starting from the center of the domain (0,0) in positive $x$-direction until point (4 mm,0). The corresponding profiles along with the crystal shape obtained after 16 hours are shown in Fig.~\ref{cc}. The $\mathit{l^2}$ norm is defined as:
\begin{equation}
{\rm E}_{\mathit{l^2}}=\sqrt{\frac{\sum_i \left( \phi_{i} - \phi_{ref,i} \right) ^2}{\sum_i \phi_{ref,i}^2}}
\label{eew}
\end{equation}
where $\phi$ represents a lower resolution and $\phi_{ref}$ denotes the highest resolution (used as reference). The errors obtained from the different simulations are illustrated in Figure~\ref{csfm}.
As observed from this plot, the numerical scheme is convergent as the error decreases with resolution. Furthermore, as expected from theoretical analyses, a second-order convergence is obtained in space.
\subsubsection{Validation against experimental data}
\begin{figure}[H]                        
\centering	
	\includegraphics[width=0.3\textwidth]{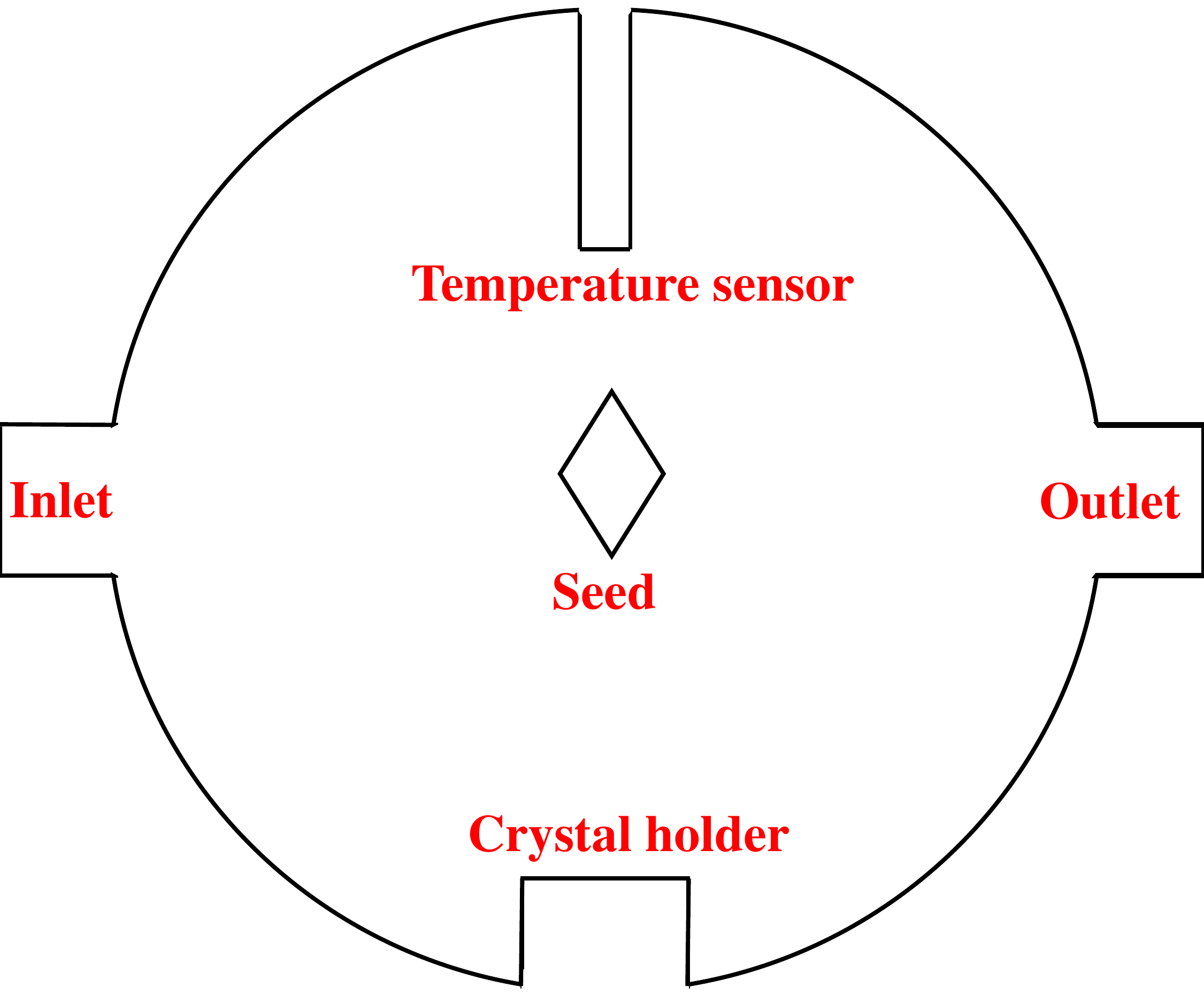}
	\caption{Reactor geometry employed for all simulations.}
	\label{bn}
\end{figure}

Next, to showcase the ability of the model to correctly reproduce the behavior of the real system, 2-D simulations are considered using the real reactor geometry. The simplification to two-dimensional simulations is justified by the fact that, in all conditions considered here, the crystal follows a platelet growth mode indicating clear separation of scales between growth in the axial and planar directions, in connection to a symmetry of the flow field~\cite{Henniges2017}. The geometry used for the simulations is shown in Fig.~\ref{bn}. First, configurations are considered where forced convection is negligible. For all experiments presented in this section the initial seed is a rhombic crystal. Two different initial supersaturations are considered, namely $U=0.06$ and $U=0.11$ for the same temperature, $T=20^\circ$C. The diffusion coefficient of mandelic acid in the aqueous solution under the conditions considered here is $D=4.273 \times 10^{-4} {\rm mm}^2/{\rm s}$~\cite{chenyakin2017diffusion} {and the other physical parameters are listed in table \ref{phy}}. Based on the excellent agreement observed in the previous section, all simulations are conducted with a spatial resolution of $\delta_r=0.025$~mm. The interface thickness is set to $W_0 = 0.05$~mm, the relaxation time to $\tau_0 = 11$~s, and the coupling coefficient $\lambda = 3$ was treated as a numerical parameter, consistently with the standard phase-field method for dendrite growth~\cite{ramirez2004phase}. At the walls of the reactor, zero-flux boundary conditions are applied to both the species and phase fields via the anti-bounce-back scheme. At the inlet a constant supersaturation is imposed. Details on the implementation can be found in~\cite{kruger2017lattice}.\\
\begin{table}[!htbp]
 \centering
 \setlength{\tabcolsep}{1.4mm}{
 \caption{{Physical parameters for a single S-ma crystal growth\cite{zhang2006evolution,suzuki2011specific,satoh1941heat}}\label{phy}}
\begin{tabular}{c c c c c}
    \hline
    Surface energy & Melting temperature & Vometric heat capacity &Latent heat &Capillary length \\ 
    $[\hbox{J} \hbox{m}^{-2}]$ & $[\hbox{K}]$ & $[\hbox{J}/\hbox{m}^3 \hbox{K}]$ & $[\hbox{J}/\hbox{m}^3]$ & $[\hbox{m}]$ \\ 
    \hline
     0.05 & 392 & 1.7 $\times 10^6$ & 6.6 $\times 10^7$ & 7.65 $\times 10^{-9}$ \\ 
    \hline
\end{tabular}}
\end{table}
Simulation results are compared to experiments and validated both qualitatively using the crystal shape, and quantitatively by comparing the growth rate.

\begin{figure}[H]                        
\centering	
	\includegraphics[width=0.4\textwidth]{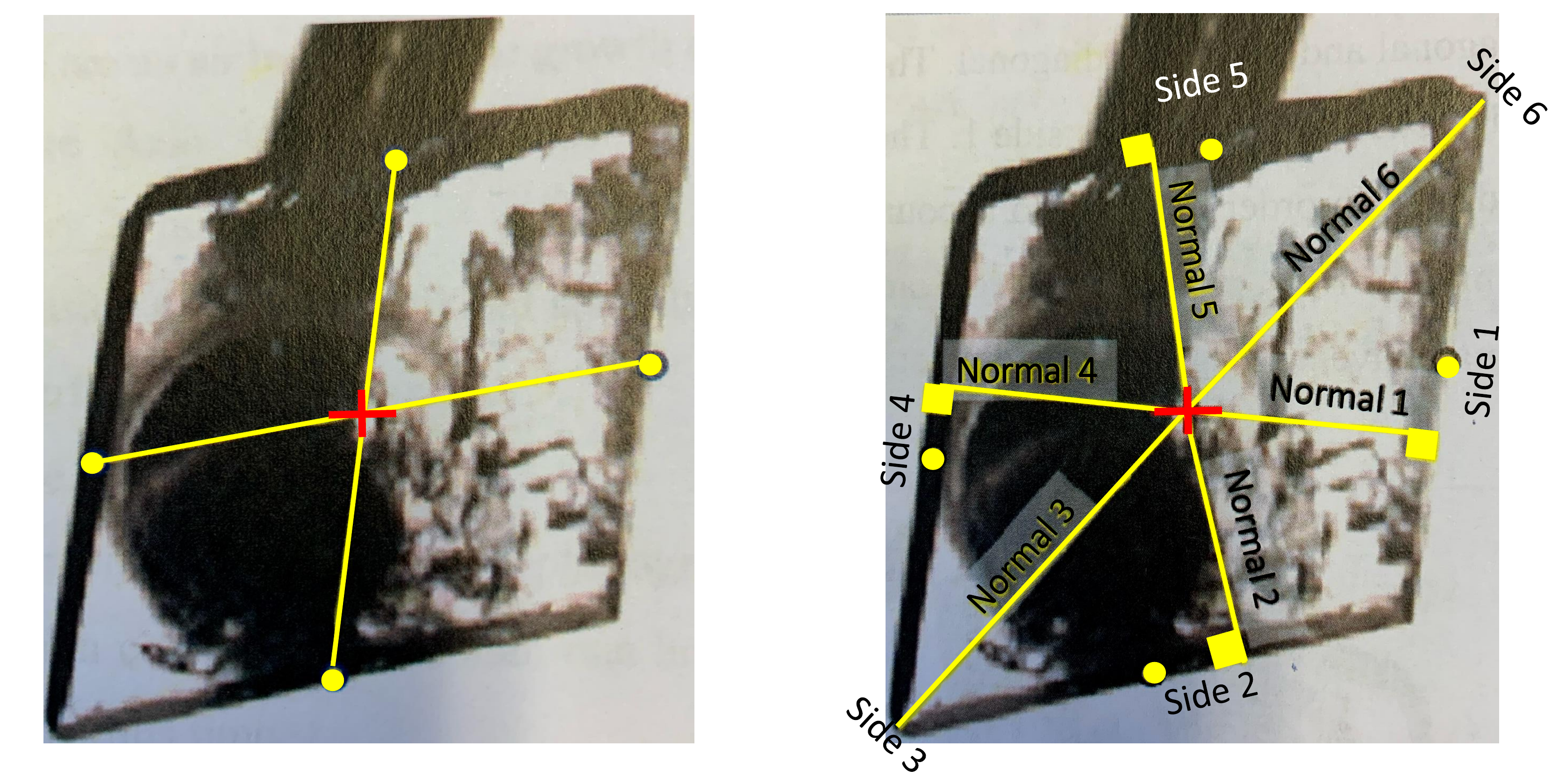}
	\caption{Method used to number the crystal sides and the associated normal directions~\cite{Juan}.}
	\label{ccf}
\end{figure}
\begin{figure}[H]                        
\centering	
	\includegraphics[width=1.0\textwidth]{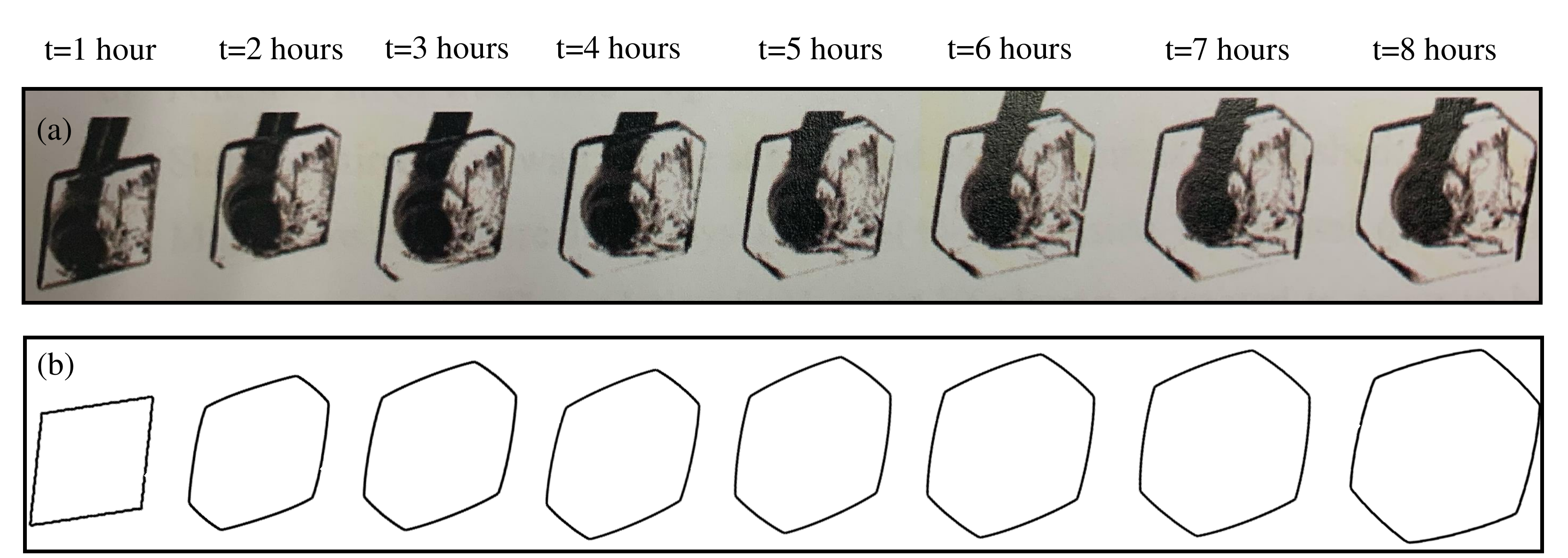}
	\caption{Contours of an (S)}-mandelic acid crystal vs time as obtained from (a) experiments~\cite{Juan} and (b) simulations. The supersaturation is $U=0.06$ in both cases. The spatial scale is the same in all images, enabling a direct comparison.
	\label{bk}
\end{figure}
\begin{figure}[H]                        
\centering	
	\includegraphics[width=0.95\textwidth]{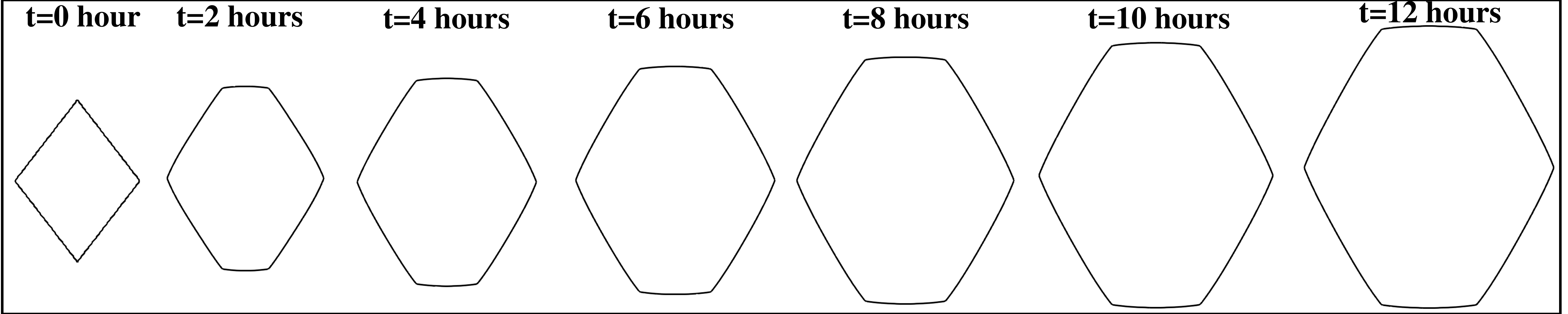}
	\caption{Contours of the (S)-mandelic acid crystal vs time as obtained from simulations for $U=0.11$.}
	\label{bak}
\end{figure}

To measure the crystal growth rate in both experiments and numerical calculations, the average length quantifying the crystal size is introduced following~\cite{Juan} as illustrated in Fig.~\ref{ccf}:
\begin{itemize}
    \item Connect opposite sides via their centers.
    \item Identify the crystal center as the intersection between those lines.
    \item Number the different sides as shown in Fig.~\ref{ccf}.
    \item Compute the lengths of the corresponding normal distance from the center identified in the previous step ($L_1$, $L_2$, $L_3$, $L_4$, $L_5$ and $L_6$). 
    \item The average normal length is simply defined as $L_{\rm avg} = (L_1 + L_2 + L_3 + L_4 + L_5 + L_6)/6$.
\end{itemize}
The experiments are systematically conducted over a period of 12 hours. The average length is computed every hour and subsequently fitted with a linear function to extract an average growth rate $G_{th}=L_{avg}/t$, where $t$ is the corresponding growth time during the crystallization process. The evolution of the crystal shape in both experiments and simulations are illustrated in Figures~\ref{bk} and~\ref{bak}.
A visual comparison regarding crystal shape and size over time for $U=0.06$ points to a good agreement between experiments and simulations. For $U=0.11$ only numerical results are shown since experimental snapshots are not available. To validate the results in a quantitative manner, the growth rates corresponding to the six different sides of the crystal for both supersaturations as obtained from experiments and simulations are compared in Table~\ref{ss}.\\
\begin{table}[!htbp]
 \centering
 \setlength{\tabcolsep}{1.4mm}{
 \caption{Comparisons between experiments and simulations for supersaturation $U = 0.06$ and $U=0.11$~\cite{Juan} \label{ss}}
\begin{tabular}{c|c|c|c|c|c|c|c|c}
    \hline
    Cases &\multicolumn{4}{c|}{Experiments}&\multicolumn{4}{c}{Simulations}\\
    \hline
   Number & \multicolumn{2}{c|}{I} & \multicolumn{2}{c|}{II} & \multicolumn{2}{c|}{I} & \multicolumn{2}{c}{II}\\ 
    \hline
    Supersaturation(\%) & \multicolumn{2}{c|}{0.06} & \multicolumn{2}{c|}{0.11}& \multicolumn{2}{c|}{0.06}& \multicolumn{2}{c}{0.11}\\ 
    \hline
    Seed perimeter (mm) & \multicolumn{2}{c|}{6.9} & \multicolumn{2}{c|}{9.5} & \multicolumn{2}{c|}{6.9} & \multicolumn{2}{c}{9.5}\\ 
    \hline
    Parameter & Slope & $R^2$ & Slope & $R^2$ & \multicolumn{2}{c|}{Slope}& \multicolumn{2}{c}{Slope}\\
    \hline
    Normal 1 & 0.07 & 0.99 & 0.14 & 0.98 &  \multicolumn{2}{c|}{0.08} & \multicolumn{2}{c}{0.14}\\
    Normal 2 & 0.09 & 0.97 & 0.14 & 0.94 & \multicolumn{2}{c|}{0.08} & \multicolumn{2}{c}{0.14}\\
    Normal 3 & 0.00 & 0.10 & 0.00 & 0.50 & \multicolumn{2}{c|}{0.01} & \multicolumn{2}{c}{0.02}\\
    Normal 4 & 0.06 & 0.94 & 0.06 & 0.78 & \multicolumn{2}{c|}{0.08} & \multicolumn{2}{c}{0.14}\\
    Normal 5 & 0.03 & 0.92 & 0.08 & 0.86& \multicolumn{2}{c|}{0.08} & \multicolumn{2}{c}{0.14}\\
    Normal 6 & 0.00 & 0.20 & 0.09 & 0.93& \multicolumn{2}{c|}{0.01} & \multicolumn{2}{c}{0.02} \\
    \hline
        Average growth rate & \multicolumn{2}{c|}{} & \multicolumn{2}{c|}{} & \multicolumn{2}{c|}{} & \multicolumn{2}{c}{}\\ 
    $G_{th}$ (mm/h) & \multicolumn{2}{c|}{0.06} & \multicolumn{2}{c|}{0.1} & \multicolumn{2}{c|}{0.057} & \multicolumn{2}{c}{0.10}\\ 
    \hline
\end{tabular}}
\end{table}
$R^2$ is the coefficient of determination shown here to characterize the reliability of the linear regression used to extract growth rates. Representing the length of side $i$ measured at time $t$ in experiments as $L_i(t)$ and the value of the linear function as $L_i'(t)$ the coefficient is computed as:
\begin{equation}
    R^2 = 1 - \frac{A_{\rm res}}{A_{\rm tot}},
\end{equation}
where the residual sum of squares is:
\begin{equation}
    A_{\rm res} = \sum_{t} {(L_i(t) - L_i'(t))}^2,
\end{equation}
and the total sum of squares is:
\begin{equation}
    A_{\rm tot} = \sum_{t} {(L_i(t) - \overline{L_i})}^2,
\end{equation}
where $\overline{L_i}$ represents the average over all data points.\\
A direct comparison of the growth rates for both values of $U$ confirms the very good agreement between experimental observations and numerical simulations. For $U=0.06$, the growth rate is numerically underpredicted by less than 6\%. At $U=0.11$, the relative difference is even reduced to 2\%. This proves the ability of the numerical model to capture the growth of (S)-mandelic acid in a pure aqueous environment. At higher supersaturation, $U=0.11$, the crystal experiences as expected a faster growth as compared to the lower supersaturation case, $U=0.06$. It is interesting to take now a closer look into the effects of supersaturation on growth rate.
\subsubsection{Impact of supersaturation on growth rate}
\begin{figure}[H]                        
\centering	
	\includegraphics[width=0.6\textwidth]{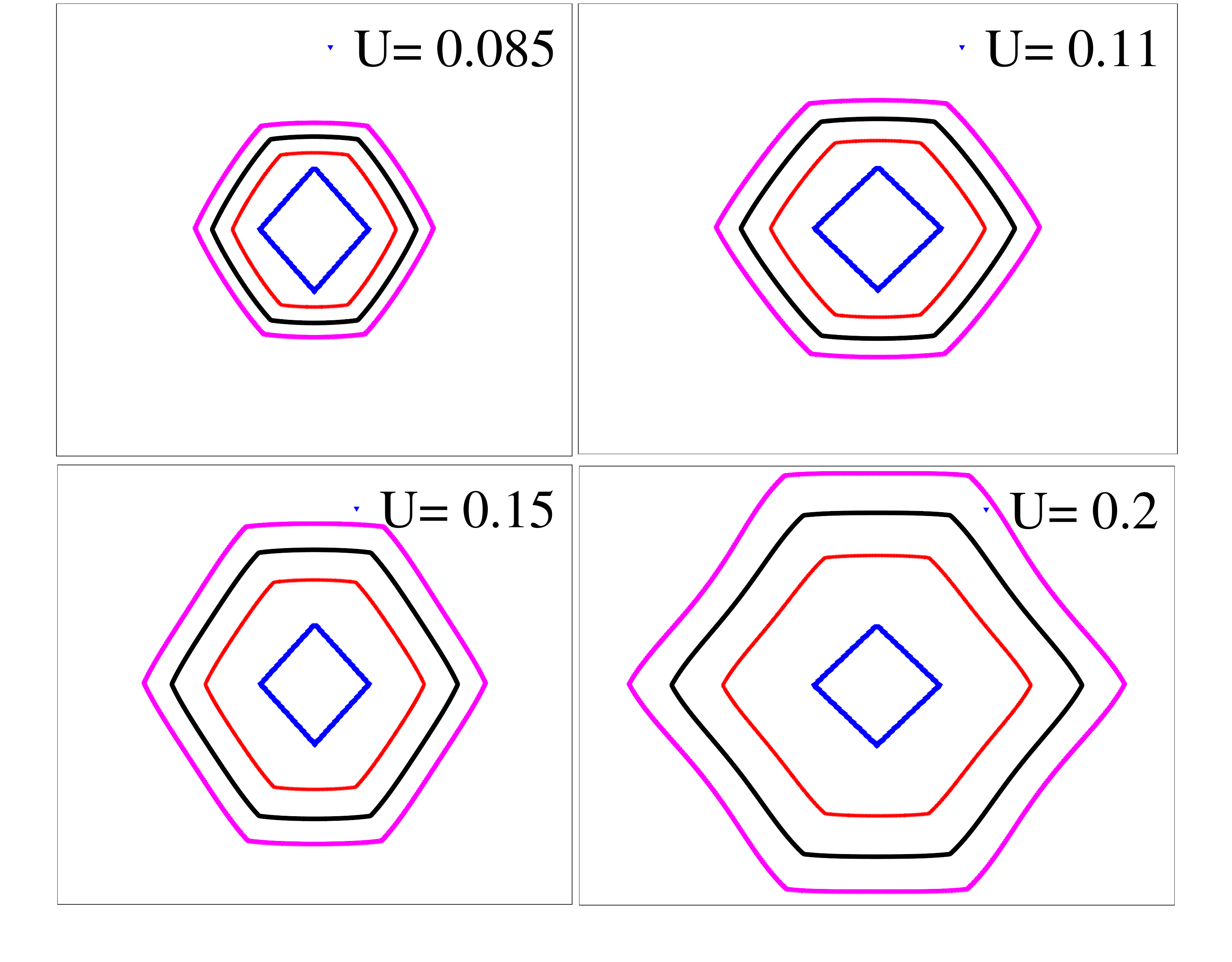}
	\caption{Boundaries of the single crystal of (S)-mandelic acid (iso-contours of $\phi=0$) at time $t$=0 (blue), 4 (red), 8 (black), 12 hours (purple) as a function of supersaturation: (top-left) $U= 0.085$, (top-right) $U= 0.11$, (bottom-left) $U= 0.15$, and (bottom-right) $U= 0.2$.}
	\label{san}
\end{figure}

In order to have a better understanding of the effect of supersaturation on crystal growth dynamics we keep a configuration similar to the previous one, but considering many more supersaturation values, $U\in\{0.06, 0.085, 0.11, 0.15, 0.2\}$. In all simulations presented in this section, the initial seed size and geometry follow that of configuration I in the previous section, see Table~\ref{ss}. The evolution of the crystal shape over time as obtained from these simulations are shown in Fig.~\ref{san}. {In Fig.~\ref{san}, the facets of the crystal start deviating from a straight line resulting from the onset of primary branching instabilities. This usually occurs when the initial value of supersaturation is sufficiently large.}
Furthermore, the data representing average growth rate are listed in Table~\ref{ljs}.
\begin{table}[!htbp]
 \centering
 \setlength{\tabcolsep}{1.4mm}{
 \caption{Numerically observed averaged crystal growth rate as function of supersaturation between $U = 0.06$ and $0.2$\label{ljs}}
\begin{tabular}{c|c|c|c|c|c}
    \hline
    Supersaturation(\%) & 0.06 & 0.085 &0.11 & 0.15 & 0.2\\ 
    \hline
    Seed perimeter (mm) & 6.9 & 6.9 & 6.9& 6.9 & 6.9\\ 
    \hline
        Average growth rate & & & & & \\ 
    $G_{th}$ (mm/h)& 0.057 & 0.0871 & 0.1136 & 0.1553 & 0.2093\\ 
    \hline
\end{tabular}}
\end{table}
As expected, higher supersaturations lead to faster crystal growth. It is worth to note that the average growth rate of the crystal can be very well approximated in the considered range as a linear function of supersaturation with slope one. Another parameter that has been observed experimentally to affect growth dynamics, especially during the early phase, is the initial seed size, better quantified via its perimeter. We will look into that effect in the next section.
\subsubsection{Impact of initial size on growth rate}
\begin{figure}[H]                        
\centering	
	\includegraphics[width=0.5\textwidth]{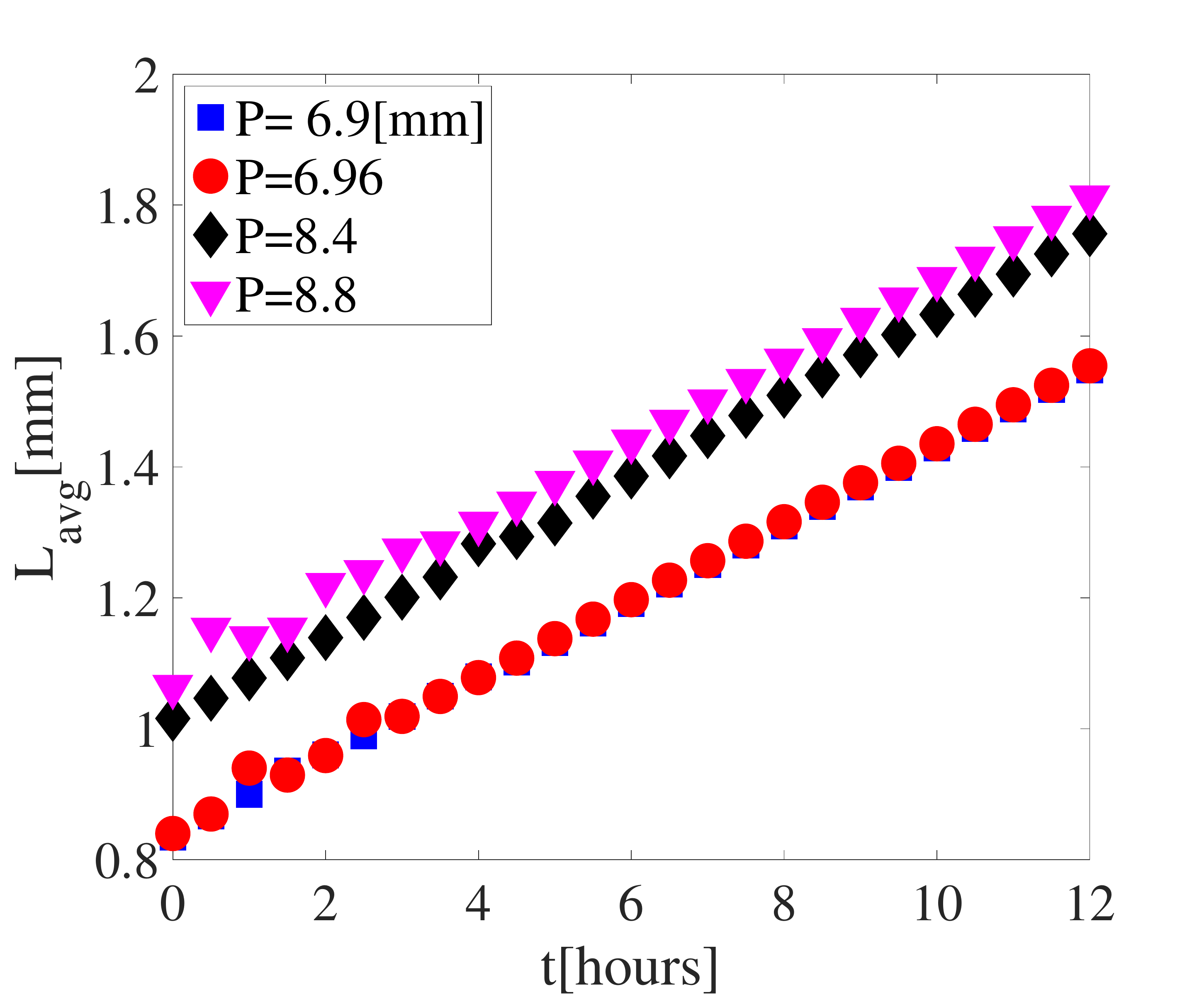}
	\caption{Numerical results concerning the effect of initial seed perimeter on the growth rate.}
	\label{ms}
\end{figure}

To quantify the effects of initial seed size, which is known to be sometimes important~\cite{srisanga2015crystal}, a configuration with supersaturation $U=0.06$ at crystallization temperature $T=20^\circ$C has been retained. Seeds with the same initial rhombic shape but different sizes have been simulated; the initial seed perimeters are $P\in\{6.9, 6.96, 8.4, 8.8\}$~mm, corresponding to available experimental data. The resulting growth rate data and parameters extracted from the experiments are listed in Table~\ref{la}.
\begin{table}[!htbp]
 \centering
  \setlength{\tabcolsep}{1.4mm}{
 \caption{Impact of initial seed size in experiments with supersaturation $U = 0.06$~\cite{Juan}\label{la}}
\begin{tabular}{c|c|c|c|c|c|c|c|c}
    \hline
    Experiment Number & \multicolumn{2}{c|}{I(1)} & \multicolumn{2}{c|}{I(2)} & \multicolumn{2}{c|}{I(3)} & \multicolumn{2}{c}{I(4)}\\ 
    \hline
    Perimeter (mm) & \multicolumn{2}{c|}{6.9}  & \multicolumn{2}{c|}{6.96} & \multicolumn{2}{c|}{8.4} & \multicolumn{2}{c}{8.8}\\ 
    \hline
    Parameter & Slope & $R^2$ & Slope & $R^2$ & Slope & $R^2$ & Slope & $R^2$ \\
    \hline
    Normal 1 & 0.07 & 0.99 & 0.09 & 0.99 & 0.09 & 0.99 & 0.10 & 0.98\\
    Normal 2 & 0.09 & 0.97 & 0.09 & 0.99 & 0.12 & 0.98 & 0.10 & 0.97 \\
    Normal 3 & 0.00 & 0.10 & 0.01 & 0.24 & 0.01 & 0.67 &  0.02 & 0.82 \\
    Normal 4 & 0.06 & 0.94 & 0.03 & 0.92 & 0.07 & 0.94 &  0.07 & 0.85 \\
    Normal 5 & 0.03 & 0.92 & 0.03 & 0.89 & 0.06 & 0.94 &  0.05 & 0.98 \\
    Normal 6 & 0.00 & 0.20 & 0.01 & 0.45 & 0.03 & 0.85  & 0.00 & 0.10 \\
    \hline
            Average growth rate& \multicolumn{2}{c|}{} & \multicolumn{2}{c|}{} & \multicolumn{2}{c|}{}  & \multicolumn{2}{c}{}\\  
    $G_{th}$ (mm/h) & \multicolumn{2}{c|}{0.06} & \multicolumn{2}{c|}{0.06} & \multicolumn{2}{c|}{0.06}  & \multicolumn{2}{c}{0.07}\\ 
    \hline
\end{tabular}}
\end{table}
Simulations with exactly the same configurations have been conducted. The corresponding results are listed in Table~\ref{lax}.
\begin{table}[!htbp]
 \centering
  \setlength{\tabcolsep}{1.4mm}{
 \caption{Impact of initial seed size in simulations with supersaturation $U = 0.06$\label{lax}}
\begin{tabular}{c|c|c|c|c}
    \hline
    Simulation Number & I(1) & I(2) & I(3) & I(4)\\ 
    \hline
    Perimeter (mm) & 6.9 &6.96 &  8.4 & 8.8\\ 
    \hline
    Parameter & Slope  & Slope & Slope  & Slope  \\
    \hline
    Normal 1 & 0.08 &  0.08 & 0.082 &0.082 \\
    Normal 2 & 0.08 & 0.08 & 0.082 &  0.082 \\
    Normal 3 & 0.01 & 0.01 & 0.012 &  0.016 \\
    Normal 4 & 0.08 &  0.08 & 0.082 & 0.082  \\
    Normal 5 & 0.08 & 0.085 & 0.085 &  0.082\\
    Normal 6 & 0.01 &  0.01 & 0.012 & 0.015 \\
    \hline
        Average growth rate & & & & \\ 
    $G_{th}$ (mm/h)& 0.0567 & 0.0575 & 0.0592& 0.0598\\ 
    \hline
\end{tabular}}
\end{table}
Both experiments and simulations, while in fair agreement with each other, point to the fact that the average growth rate is only slightly affected by the initial seed size. The effect appears to be somewhat stronger for a larger initial seed. The growth behavior over time is illustrated in Fig.~\ref{ms}. The results for initial perimeters $6.9$ and $6.96$ mm cannot be differentiated visually. For a larger initial seed, the differences between $8.4$ and $8.8$ mm can be recognized, but only a minute increase in average growth rate can be recognized. It can be concluded that, compared to supersaturation, the influence of initial seed size is minor. Still, a larger initial seed corresponds to a slightly increased growth rate.
All the previous results have been obtained while neglecting any convection effect around the crystal. However, it is known that forced convection might lead to asymmetric crystal growth. This will be explored next.
\subsection{Ventilation effects}
\begin{figure}[H]                  
\centering	
	\includegraphics[width=0.7\textwidth]{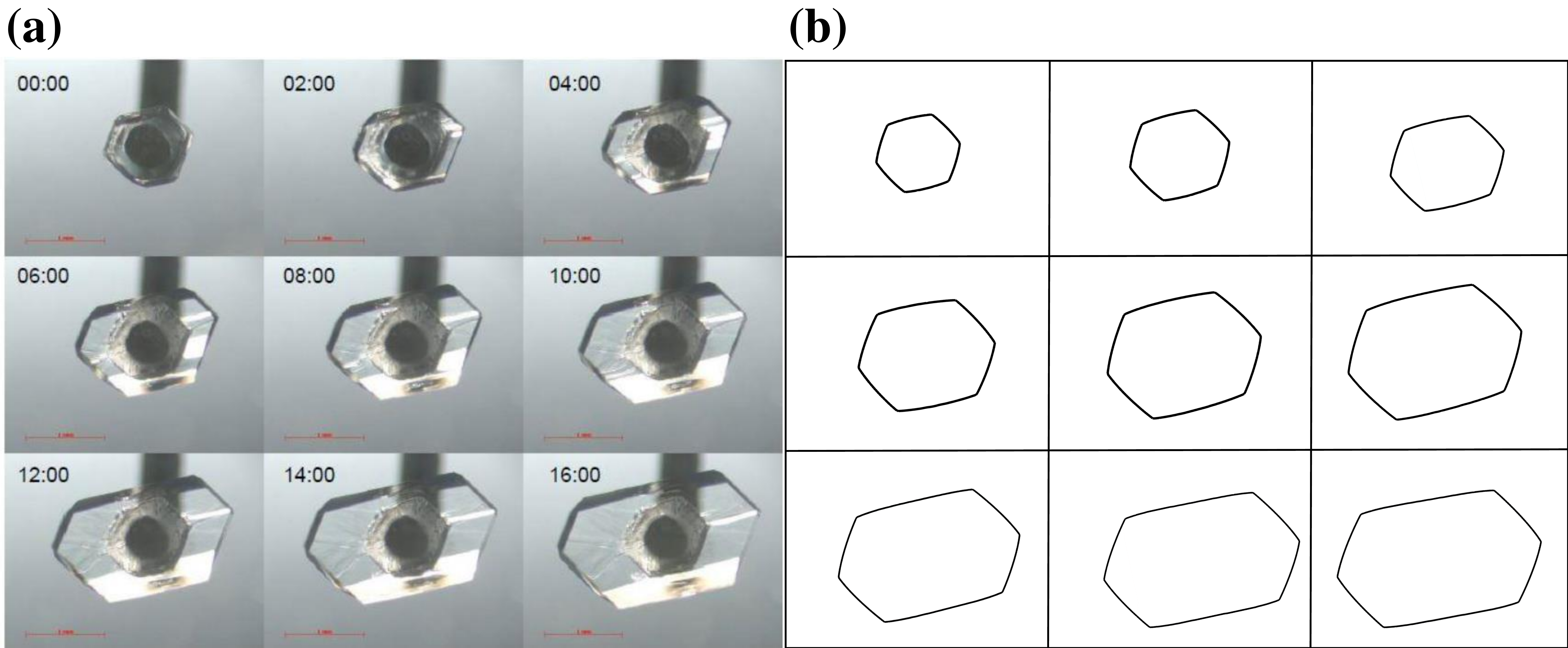}	
	\caption{Morphologies of (S)-mandelic acid crystal captured by (a) camera during the experiments~\cite{Linzhu}; (b) simulations.}
	\label{ta}
\end{figure}

\begin{figure}[H]  
\centering	
	\includegraphics[width=0.8\textwidth]{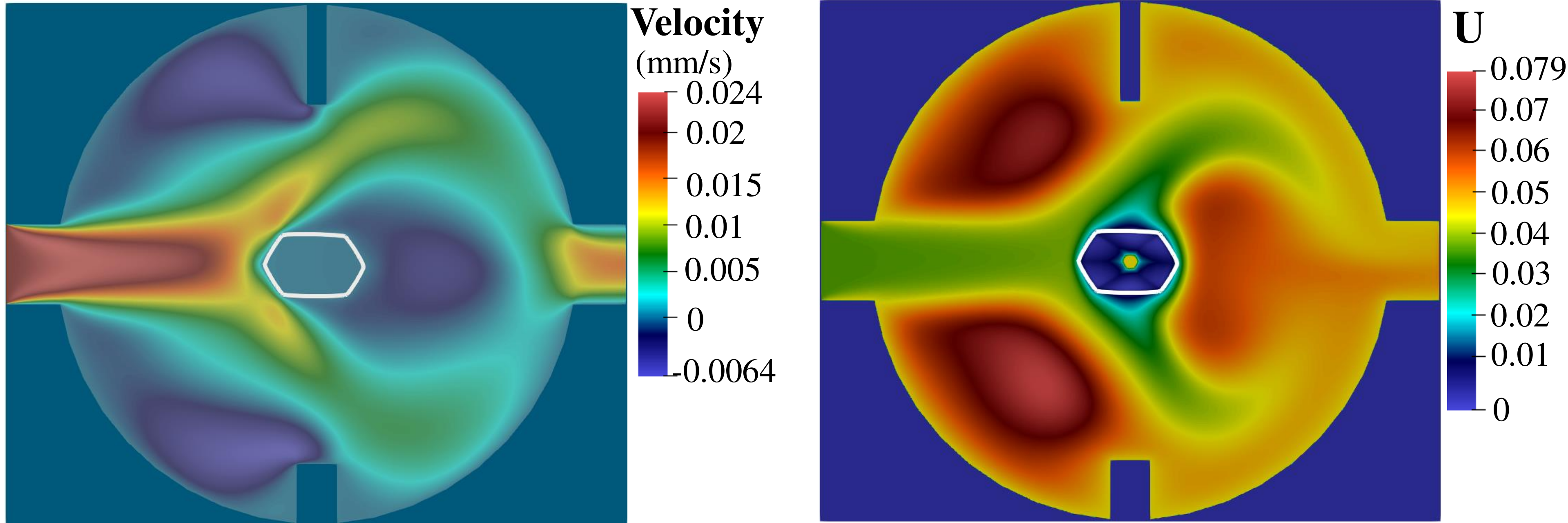}	
	\caption{Non-symmetric growth of a (S)-mandelic acid crystal taking into account convection as obtained from the simulation for $U=0.045$ after 16 hours. Flow is from left to right in the reactor.}
	\label{ai}
\end{figure}

\begin{figure}[H]  
\centering	
	\includegraphics[width=0.8\textwidth]{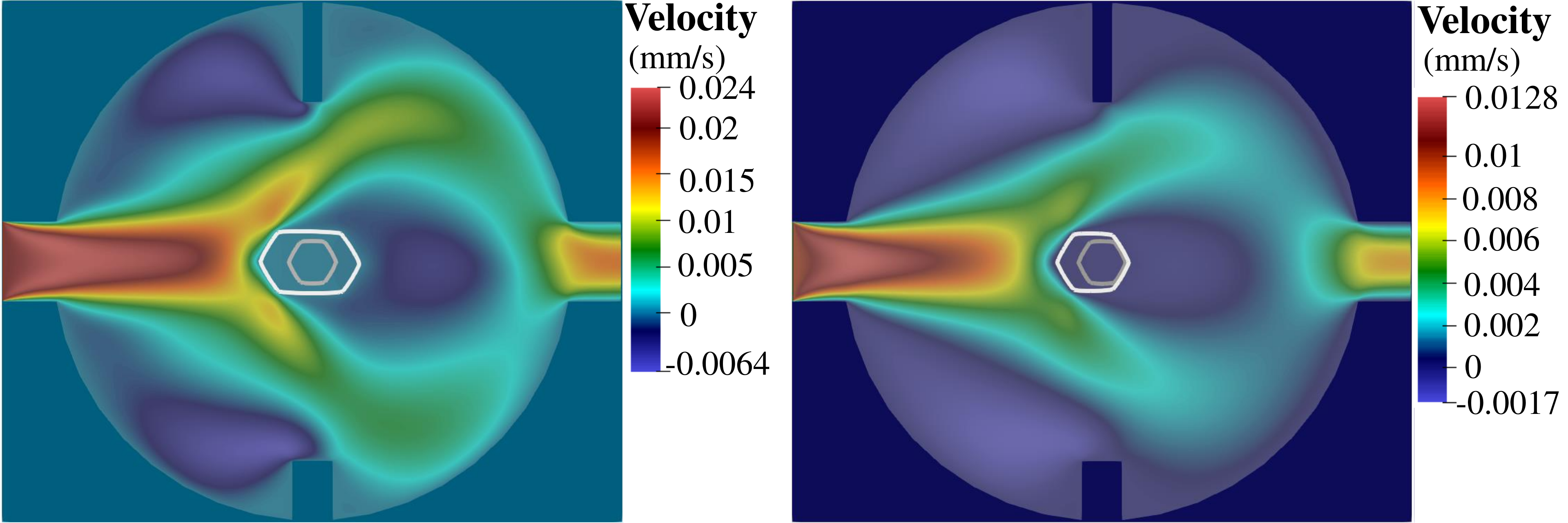}	
	\caption{Convection effects on (S)-mandelic acid crystal growth after 10 hours for $U=0.045$ and two different Reynolds numbers. Left side: Re $= 17.2$; Right side: Re $= 8.6$}. The white line represents the crystal boundary taking into account the flow (ventilation effect), while the grey line shows the same results in the absence of any inflow.
	\label{aib}
\end{figure}

\begin{figure}[H]  
\centering	
	\includegraphics[width=0.5\textwidth]{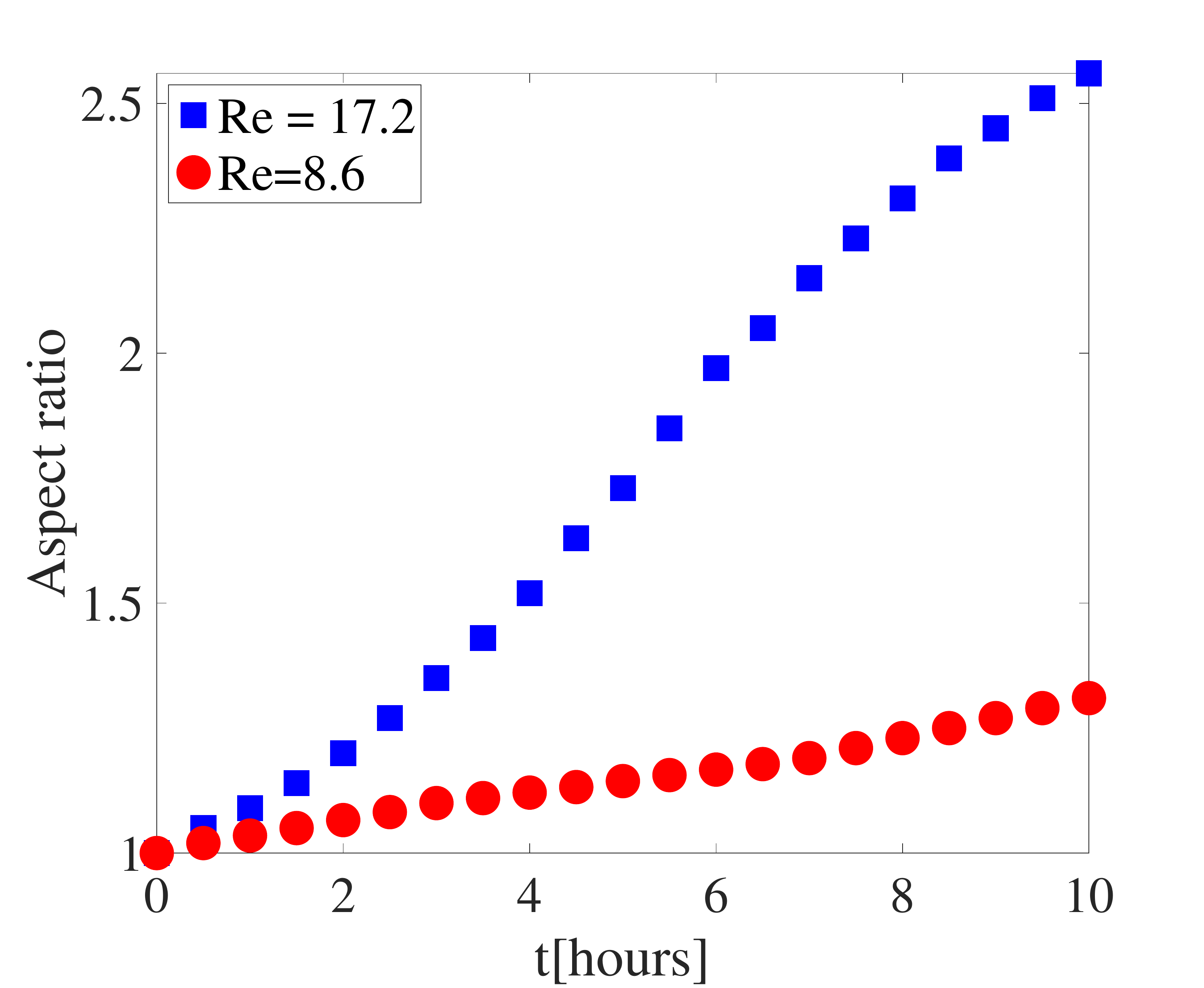}	
	\caption{Evolution of the aspect ratio vs time for Reynolds numbers Re$=8.6$ and $17.2$.}
	\label{aicb}
\end{figure}

\begin{figure}[H]  
\centering	
	\includegraphics[width=0.7\textwidth]{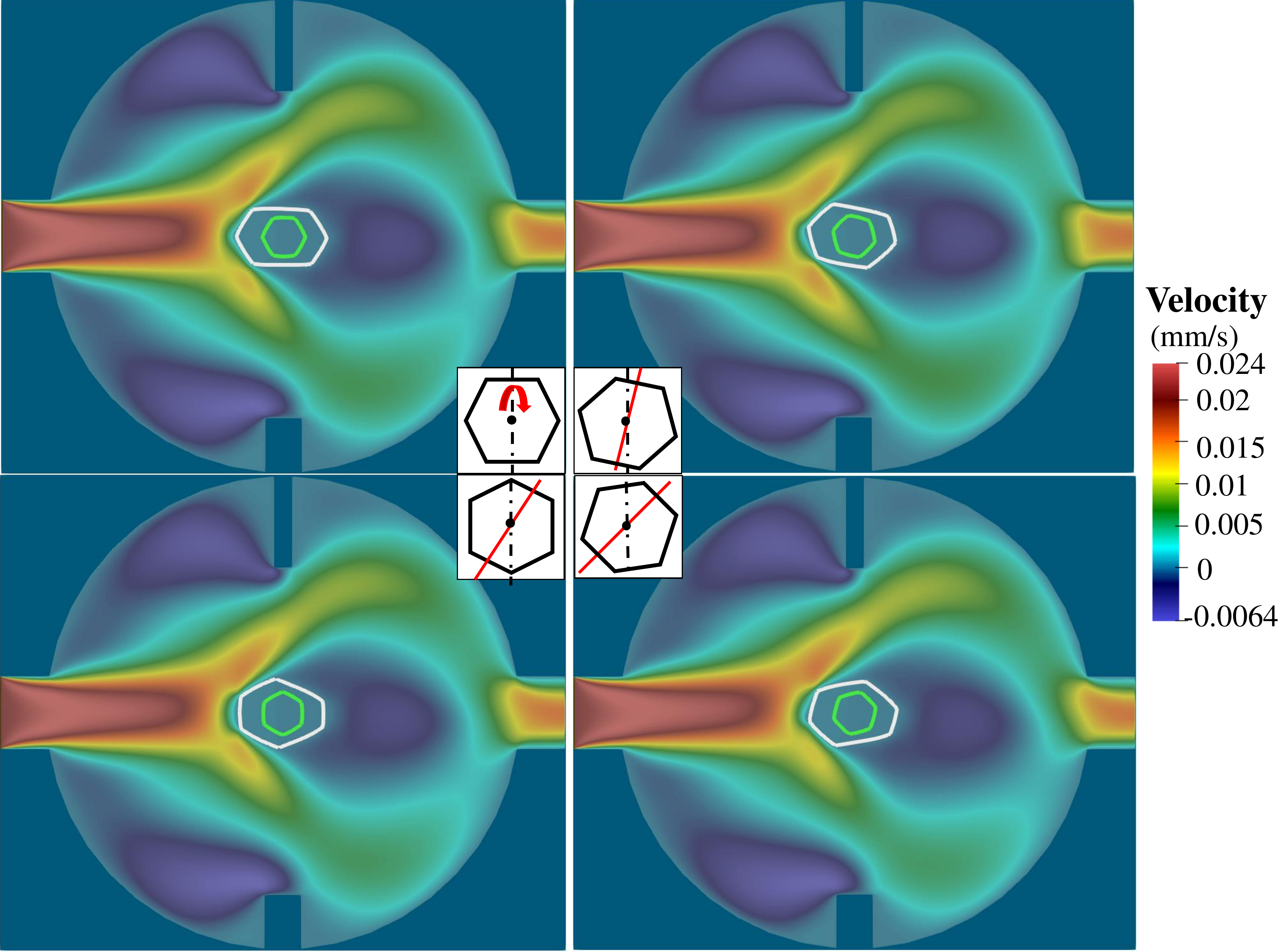}	
	\caption{Effect of initial seed orientation on (S)-mandelic acid crystal growth after 10 hours for $U=0.045$. Top left: without rotation; Top right: initial rotation of  $\frac{\pi}{12}$ (clockwise rotation); Bottom left; initial rotation of $\frac{\pi}{6}$; Bottom right: initial rotation of $\frac{\pi}{4}$. The white line represents the crystal boundary taking into account the flow, while the green line shows the same results in the absence of any inflow.}
	\label{aia}
\end{figure}

\begin{figure}[H] 
\centering	
	\includegraphics[width=0.35\textwidth]{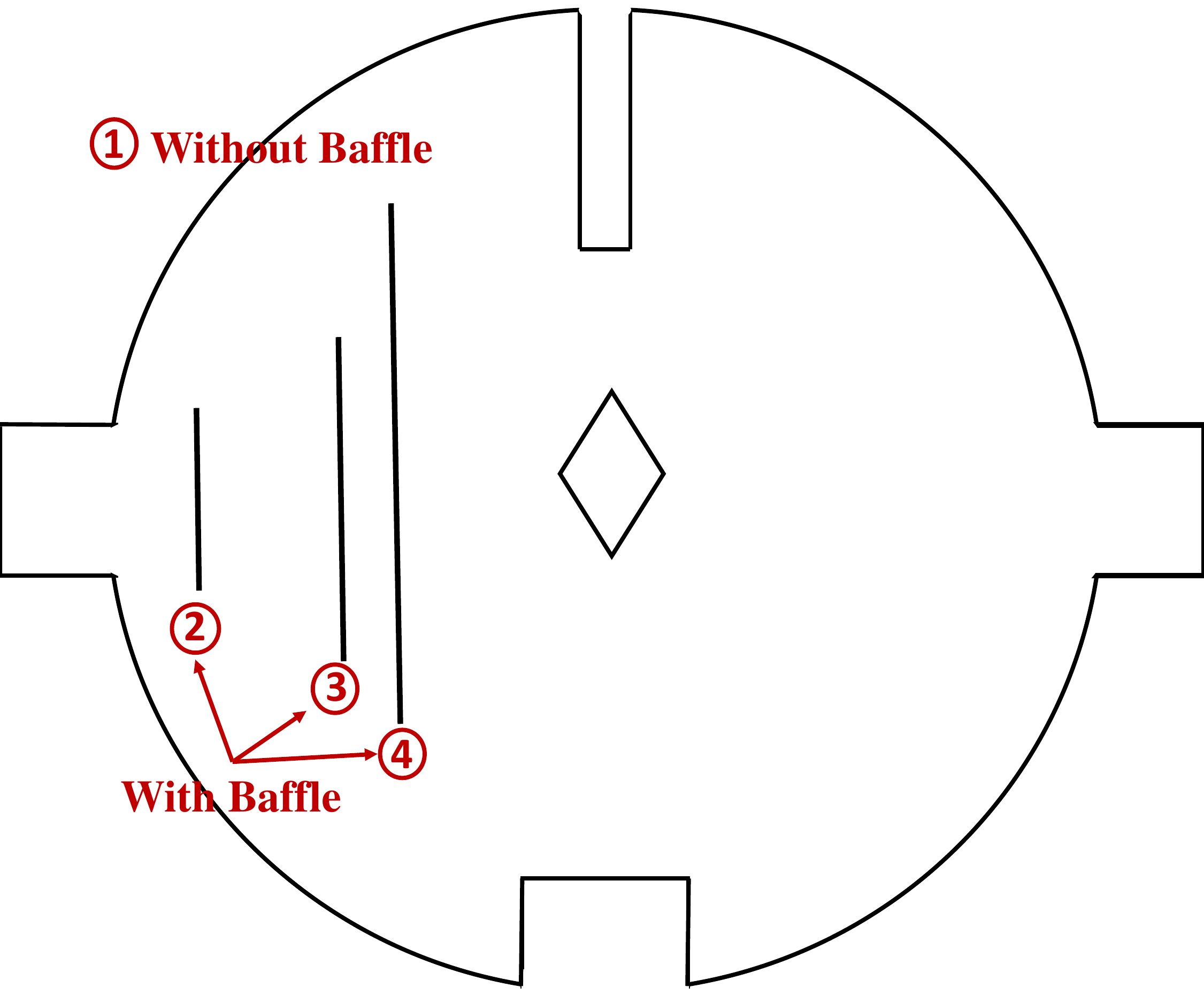}	
	\caption{Proposed modifications of the geometry of the single-crystal growth cell reactor including a baffle (three different possible configurations).}
	\label{aih}
\end{figure}

\begin{figure}[H] 
\centering	
	\includegraphics[width=0.9\textwidth]{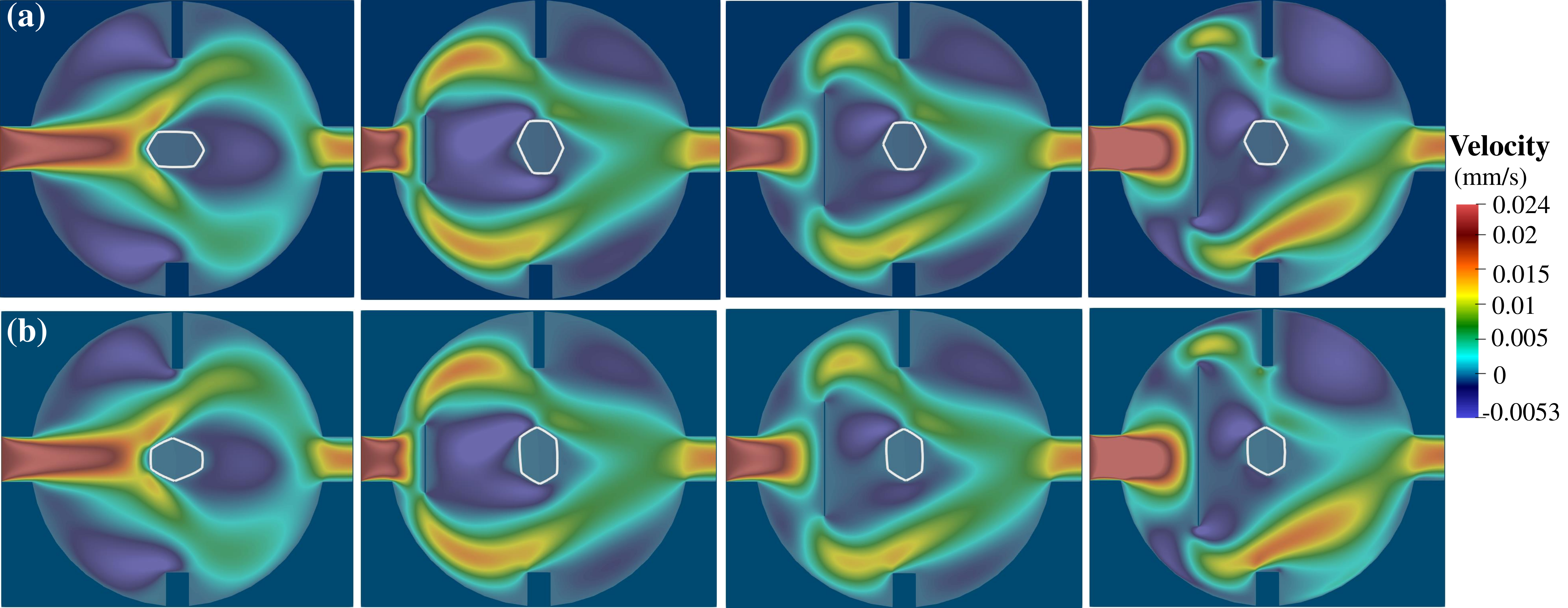}	
	\caption{Numerical prediction for the growth on (S)-mandelic acid crystal after 10 hours for $U=0.045$ and Re$= 17.2$. First column: original configuration, without baffle; second column: baffle configuration 2; third column:  baffle configuration 3; fourth column:  baffle configuration 4. Top line (a): without any rotation of the initial seed; Bottom line (b): with initial rotation of the seed by $\pi$/6.}
	\label{aif}
\end{figure}

\begin{figure}[H] 
\centering	
	\includegraphics[width=0.9\textwidth]{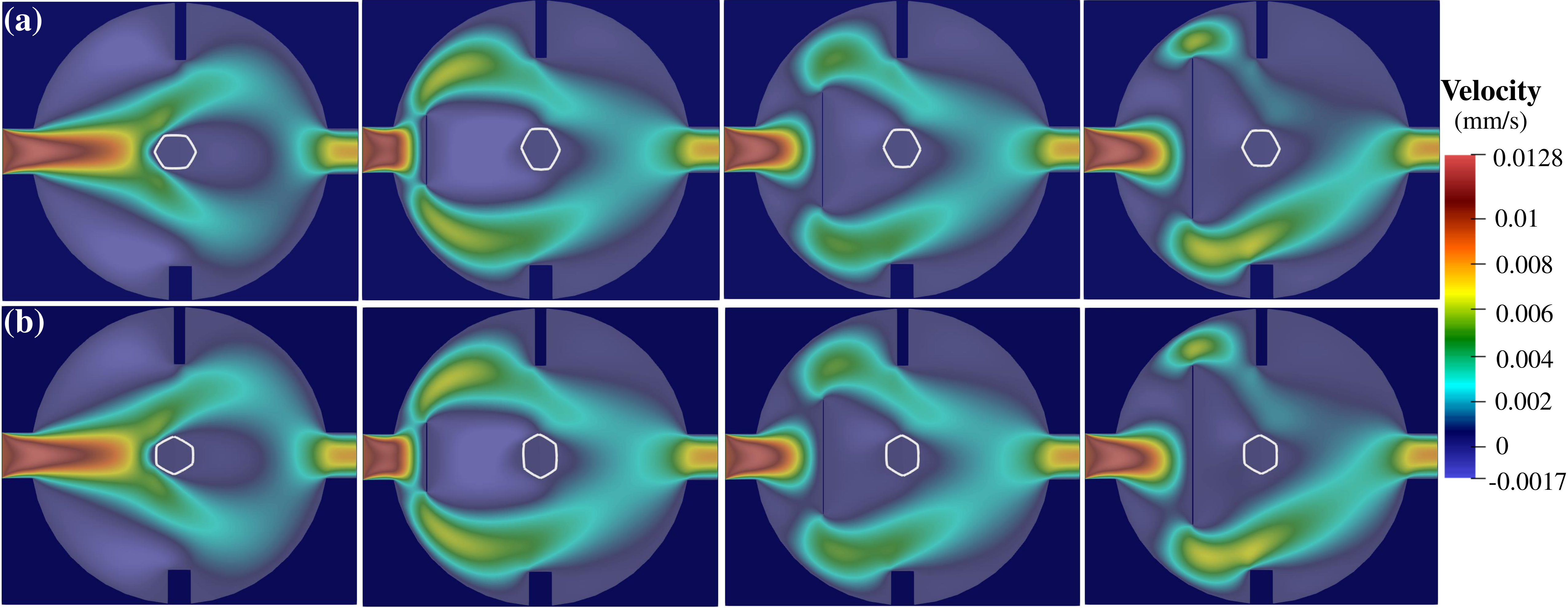}	
	\caption{Numerical prediction for the growth on (S)-mandelic acid crystal after 10 hours for $U=0.045$ and Re$= 8.6$. First column: original configuration, without baffle; second column: baffle configuration 2; third column:  baffle configuration 3; fourth column:  baffle configuration 4. Top line (a): without any rotation of the initial seed; Bottom line (b): with initial rotation of the seed by $\pi$/6.}
	\label{aig}
\end{figure}

In the real single-crystal reactor the incoming flow of (S)-mandelic acid in water might have a large impact on crystal growth rate and shape development. The aim of the present section is to check and quantify this point.
\paragraph{Validation in presence of convection}
First, we validate the numerical model against available experimental data taking into account the real inflow conditions used in the reactor cell. For this purpose, and following the experimental settings~\cite{Linzhu}, a hexagonal seed of perimeter $P = 3.7$~mm is used. The initial supersaturation is $U=0.045$, the Reynold number Re = 17.2. Here, we keep the same Reynold number as the experiment~\cite{Linzhu}. The results, represented by the crystal shape, are compared over time via snapshots taken every two hours over an overall growth period of 16 hours, as shown in Fig.~\ref{ta}.
It is observed that the evolution of the crystal shape over time as obtained from both experiment and simulation are in good agreement with each other; they both point to a non-symmetrical growth. The constant inflow of a solution with a higher concentration hitting the inlet-facing sides of the crystal subjects them to noticeably larger gradients at the interface, as compared to the other, leeward sides; this induces lower adsorption rates at the latter. As a result the (S)-mandelic crystal grows faster on the sides facing the inflow, leading to a steady increase of the aspect ratio, defined as the ratio of the horizontal size of the crystal to its vertical size. This is clearly visible in Fig.~\ref{ai} where both the velocity and supersaturation fields at $t=$16 hours are shown. 
To further illustrate the effects of hydrodynamics on the crystal habit, the effect of the Reynolds number and of the initial orientation of the seed are considered numerically.
\paragraph{Effect of Reynolds number}
For this purpose, the supersaturation is kept constant at $U=0.045$. The Reynolds number is the most important non-dimensional parameter of fluid dynamics, comparing quantitatively convective effects to dissipation by viscosity. It is defined as:
\begin{equation}
    {\rm Re} = \frac{u_{\rm in} P}{\nu_f},
\end{equation}
where $P$ is the initial seed perimeter. Two different Reynolds numbers, i.e. Re$=8.6$ and $17.2$, are considered. The obtained crystal shape and velocity fields are shown in Fig.~\ref{aib}.
As expected, higher inlet velocities hitting the inflow-facing sides of the crystal result in a faster growth in that direction; the resulting asymmetry becomes more marked, leading to elongated crystals in the horizontal direction for the present setup. This is particularly clear looking at Fig.~\ref{aicb}. After 10 hours, the aspect ratio for Re$=17.2$ is already more than twice as large than for Re$=8.6$.
\paragraph{Effect of the initial orientation of the seed}
To show the effect of seed orientation, four simulations have been carried out with the same Reynolds number, Re$=17.2$, but with different initial orientations, namely $\theta\in\{0, \frac{\pi}{12}, \frac{\pi}{6}, \frac{\pi}{4}\}$. This choice of tilt is motivated by the six-fold symmetry of the crystal natural habit, so that only the 0-$\pi/3$ range is relevant. The obtained crystal habits after 10 hours along with the corresponding velocity fields are illustrated in Fig.~\ref{aia}.  
It is seen that the initial orientation not only affects the symmetry of the crystal, but also its average growth rate. Taking into account convective effects and initial seed orientation, the crystal habits become highly asymmetrical. It is also observed that a slight initial rotation in clockwise direction can result in a final habit showing preferential counter-clockwise orientation, due to a strong interaction with the convective flow field.
\paragraph{Improving symmetrical growth in presence of convection using a} baffle
As seen from the previous simulations, the overall shape of the crystal varies considerably as function of the Reynolds number. It was mentioned earlier that the regularity of the crystal shape is a property of high interest regarding the final products performance. Therefore, it is desirable to find a simple geometrical modification to the single-crystal growth cell leading to isotropic growth rates and a desired final aspect ratio. For this purpose, a simple flat baffle has been placed in the simulation directly in front of the inlet in order to prevent a direct impact of the incoming flow onto the growing seed. Three different configurations (different position, different size) of the baffles have been compared. The resulting configurations are illustrated in Fig.~\ref{aih}; configuration 1 is the original case, without any baffle.
To check the robustness of the proposed modification with the three different baffles (plus the original case), two different Reynolds numbers (Re$=8.6$ or $17.2$), and two different initial seed orientations ($\theta$=0 or $\pi/6$) have been considered, making up for a total of ($4\times 2\times 2=$) 16 different cases. All numerical results after 10 hours of growth are shown in Figs.~\ref{aif} (for Re=$17.2$) and \ref{aig} (for Re$=8.6$).

To quantify the effects of the baffles on the quality of the crystal, a quality parameter is defined as $Q=\max(L_i)/\min(L_i)$ where index $i\in\{0,\dots,5\}$ covers the length of all sides of the resulting crystal. Parameter $Q$ quantifies non-isotropic growth, with $Q=1$ (the minimum value) corresponding to a perfectly isotropic growth, while an increasing value of $Q$ corresponds to growing non-isotropy. The values of crystal quality as obtained from all simulations after 10 hours of growth are listed in Table~\ref{zz}.
\begin{table}[!htbp]
 \centering
  \setlength{\tabcolsep}{1.4mm}{
 \caption{Impact of the baffles for $U = 0.06$ for two different Reynolds numbers and seed orientations\label{zz} in Fig.~\ref{aih}}
\begin{tabular}{c|c|c|c|c}
    \hline

    Q & Re=8.6; tilt=0 &Re=8.6; tilt=$\pi$/6 & Re=17.2; tilt=0 &  Re=17.2; tilt=$\pi$/6\\
    \hline
    No Baffle & 1.28 & 1.23 & 1.625 & 1.59\\
    Baffle 1 & 1.22 & 1.21 & 1.41 &  1.39  \\
    Baffle 2 & 1.13 & 1.16 & 1.24 & 1.21 \\
    Baffle 3& 1.05 & 1.07 & 1.14 & 1.12 \\
    \hline
\end{tabular}}
\end{table}
From Table~\ref{zz} it is clearly observed that, while all baffles contribute to reducing asymmetrical growth, the larger one, i.e. baffle 3, leads to the best crystal quality in terms of symmetry for all considered conditions. It successfully reduces deviations from perfect symmetry by about 20\% for Re$=8.6$ and 30\% for Re$=17.2$. Since the complexity of the experimental setup would not be significantly increased by adding a baffle, such a modification is recommended for further studies.
\section{Conclusions and perspectives}
In this work, a numerical model based on lattice Boltzmann has been developed and validated to describe crystal growth. It has been shown to correctly capture the dynamics of (S)-mandelic acid crystal growth. The numerical simulations were compared to experimental data from a single-crystal growth reactor and are in very good agreement. The model was then used to investigate the effects of important parameters such as supersaturation and initial seed size on growth dynamics. It was depicted that higher supersaturation levels lead to much faster growth rates; the impact of a larger initial seed crystal is by far not as strong, but increases slightly the growth rate as well.\\
It was also demonstrated that hydrodynamics can have pronounced effects on both average growth rate and habit, and may lead to a clear rupture of symmetry. The evolution of the crystal habit was shown to change significantly with the Reynolds number, but also with the initial orientation of the seed with respect to the incoming flow. Finally, a simple modification of the reactor geometry was proposed to minimize non-symmetrical growth. This will be tested in later experiments.\\
While the model was successfully applied in the present study to pure (S)-mandelic acid crystal growth under isothermal conditions, in the future, these simulations will be extended to more complex situations involving a mixture of both (S)- and (R)-mandelic acid and taking into account temperature changes.
\section*{Acknowledgments}
Q.T. would like to acknowledge the financial support by the EU-program ERDF (European Regional Development Fund) within the Research Center for Dynamic Systems (CDS). S.A.H. acknowledges the financial support of the Deutsche Forschungsgemeinschaft (DFG, German Research Foundation) in TRR 287 (Project-ID 422037413).
The authors gratefully acknowledge the computing time granted by the Universit\"at Stuttgart-H\"ochstleistungsrechenzentrum Stuttgart (HLRS); all calculations for this publication were conducted with computing resources provided under project number 44216.\\
\bibliographystyle{plain}
\bibliography{ref.bib}
\end{document}